\documentclass[12pt,floatfix,eqsecnum]{revtex4}%
\usepackage{amsmath}
\usepackage{graphicx}
\usepackage{multirow}
\usepackage{color}
\usepackage{amssymb}
\usepackage{amsfonts}%
\begin{document}
\title{A Rational Method for Probing Macromolecules Dissociation: The Antibody-Hapten System}
\author{Elsa S. Henriques}
\email{henriques@fias.uni-frankfurt.de}
\affiliation{Frankfurt Institute for Advanced Studies, Max-von-Laue-Strasse 1, D-60438
Frankfurt am Main, Germany}
\author{Andrey V. Solov'yov}
\email{solovyov@fias.uni-frankfurt.de}
\affiliation{Frankfurt Institute for Advanced Studies, Max-von-Laue-Strasse 1, D-60438
Frankfurt am Main, Germany}
\date{\today}

\begin{abstract}
The unbinding process of a protein-ligand complex of major biological interest
was investigated by means of a computational approach at atomistic classical
mechanical level. An energy minimisation-based technique was used to determine
the dissociation paths of the system by probing only a relevant set of
generalized coordinates. The complex problem was reduced to a low-dimensional
scanning along a selected distance between the protein and the ligand.
Orientational coordinates of the escaping fragment (the ligand) were also
assessed in order to further characterise the unbinding. Solvent effects were
accounted for by means of the Poisson--Boltzmann continuum model. The
corresponding dissociation time was derived from the calculated barrier
height, in compliance with the experimentally reported Arrhenius-like
behaviour. The computed results are in good agreement with the available
experimental data.

\end{abstract}

\pacs{}
\keywords{}\maketitle

\section{Introduction}

Biological processes are driven by interactions between the molecular
components of cellular machinery, commonly between proteins and their target
molecules (generically termed ligands). Most of these processes portray a
cascade of protein-ligand association/dissociation events, and thus, knowledge
and control of their energetics and kinetics is of key importance in molecular
biology, proteomics, clinical diagnosis, and therapeutic research, to name a few.

Protein-ligand dissociation is, in essence, a fragmentation of complex
multi-atomic aggregates. Many-body aggregates are very ubiquous in Nature, and
have been the object of extensive experimental and theoretical studies in a
wide range of natural science research fields: examples range from nuclear
fission to atomic clusters fragmentation to dissociation of insulin from its
receptor on the cell membrane, etc. A vast amount of data has now been
accumulated, but there is still a need for an efficient and physically sound
theoretical approach that could possibly rationalize these data and make
insightful predictions, the applicability of one such approach being obvious.
A first step is to try and identify the common features underlying
dissociation events of different nature.

Clustering and fragmentation processes in nuclear and atomic cluster physics
have already been found to possess many features in common (for a
comprehensive review see ref. \cite{ASolovyovIB01}). The emerging key idea is
that those processes can be successfully described in terms of a few
collective coordinates that define the overall geometry configuration of the
escaping and parent fragments \cite{ASolovyovIB01,LyalinA01}. The same basic
concept also holds for similar processes in more complex systems, like the
fragmentation of a dipeptide \cite{ISolovyovA01}. On the basis of this
principle, the present paper addresses the dissociation process of an
aggregate of higher complexity, a biological protein-ligand adduct (often
referred as a complex).

\begin{figure*}[t]
\centering
\includegraphics[scale=0.80,clip]
{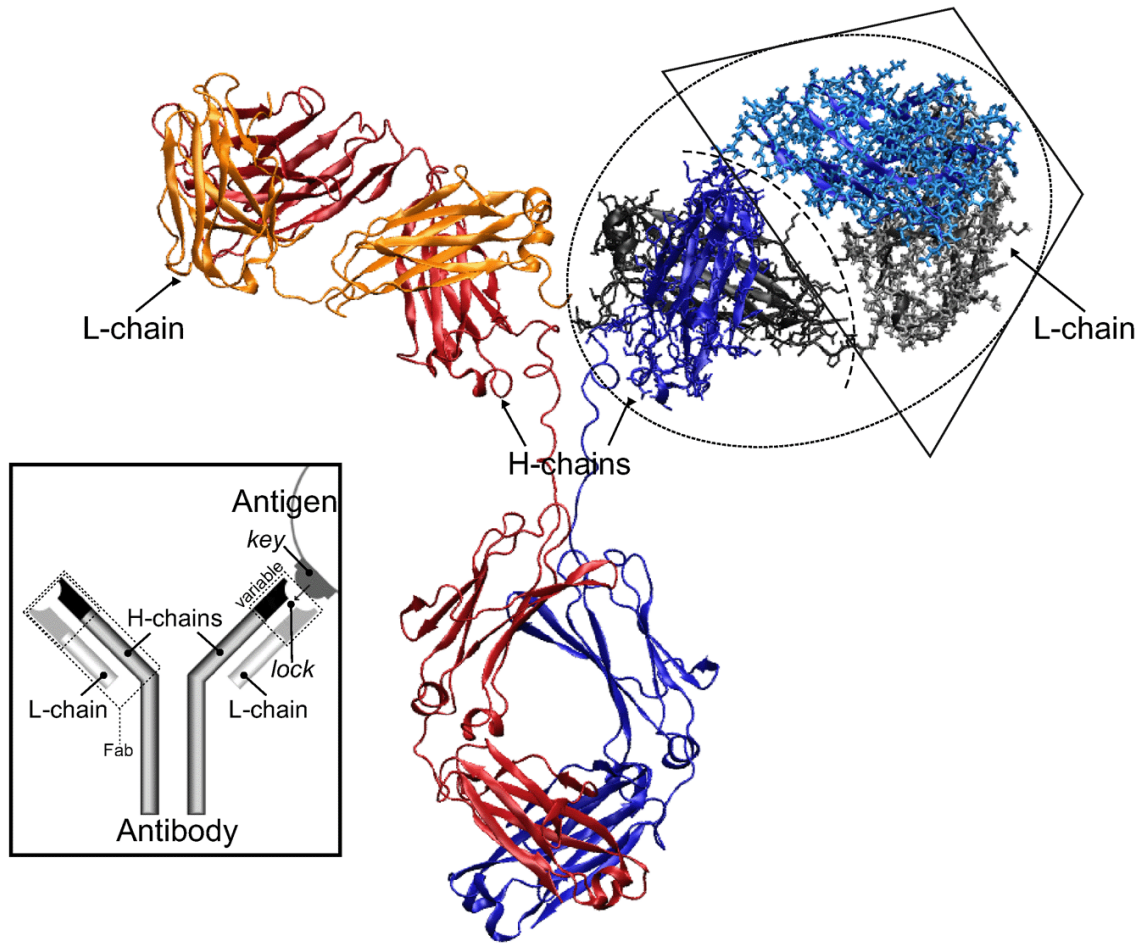}
\caption{Overall ribbon representation of a complete AB structure. The two
pairs of heavy chains are depicted in red and blue, and the corresponding
light chains in yellow and grey. The dashed ellipse highlights one of the
branches that bind to the antigen (the so called Fab, after fragment binding antigen),
in an all-atom representation; the trapezoidal region puts in evidence
the Fab variable domains (with added hydrogens), and the dashed arc
illustrates the chains' cleavage sections for these variable domains to be
detached. A simplified scheme of AB-AG binding is presented in the inset.}
\label{F1}
\end{figure*}

A most remarkable protein-ligand system is the antibody-antigen one, which is
involved in a fundamental recognition process during the body immune response.
This response is triggered by foreigner molecules -- the antigens
(\underline{anti}body \underline{gen}erator, AG). One key mechanism whereby
the immune system recognizes and targets them for destruction is by releasing
antibodies (\underline{anti}-foreign \underline{body}, AB) \cite{WedemayerA01}.
ABs are very large proteins, and the human body has a potential repertoire
of 2.5$\times$10$^{11}$ different ones. Yet, they all feature a basic scaffold:
they consist of two identical ``light'' (L) and ``heavy'' (H) chains of amino acids
entangled in a Y-shape fold as shown in Figure \ref{F1}. Each tip of the Y
branches displays a distinctive variable region, \textit{i.e.}, the specific
``lock'' for which the target AG has the ``key'' (see the schematic inset in
Figure \ref{F1}); the two tips are identical for each AB. The ``key'' region
of the AG can be a small protein fragment or a
hapten. A hapten is a low molecular weight compound originally attached to
some carrier protein, that will also trigger the release of ABs. Upon exposure
to a particular AG, a set of ABs is refined to target it, \textit{via} a mutation
process \cite{ManserA01,LehningerB01}. The mutations occur in the referred variable
region (hence it is called ``variable'').
Along a maturation series, the increase in affinity strongly correlates with
an increase in the corresponding AB-AG dissociation times, $\tau$
\cite{WedemayerA01,SchwesingerA01,FooteA01,JimenezA01}. Usually, $\tau$ is
expressed in terms of the rate of spontaneous dissociation, $k_\mathrm{off} = 1/\tau$.

Not surprisingly, much effort has been devoted to the determination of those
$k_\mathrm{off}$ values, with some of the most innovative experiments involving sensitive
micromanipulation techniques like atomic force microscopy (AFM) and other
force probe procedures to measure AB-AG binding forces
\cite{WedemayerA01,HinterdorferA01,DammerA01,AllenA01,SulchekA01}. Some
further insight into the molecular structure, interactions and unbinding
pathways underlying such single molecule experiments has been gained from
computer simulations using  ``force probe''
molecular dynamics (FPMD) \cite{GrubmullerIB01}. However, the question
arises of to what extent the measured unbinding force in the mechanically
speeded up process of pulling out the ligand relates to the thermodynamic or
kinetic parameters describing the spontaneous dissociation. The later arises
in the minute time scale \cite{FooteA01} in contrast to the time scales of AFM
(millisecond) and FPMD (nanosecond). There is also the matter of across which
pathway is unbinding being forced.

In the absence of a pulling force, one regains the spontaneous (natural) mode
of AB-AG dissociation, a thermally activated barrier-crossing along a
preferential path in a multidimensional energy landscape. The contributing
activated states (which determine $k_\mathrm{off}$) may well be described in terms of
a few collective coordinates, in close analogy to other studied fragmentation
processes \cite{ASolovyovIB01,LyalinA01,ISolovyovA01}. Within this context, it
is reasonable to constrain the many other degrees of freedom that only
contribute to the negligible fine structure of the energy landscape. This is a
rational approach to probe the unbinding of a complex biological system like
the AB-AG one, in order to calculate the corresponding energetic barrier and
derive $k_\mathrm{off}$ from it.

Starting with an experimentally well studied AB-AG complex, an
anti-fluorescein one (\textit{vide infra}), here we describe a computational
approach at molecular (atomistic) level to explore its preferential unbinding
pathways by probing only a few relevant degrees of freedom. A detailed
analysis of its dissociation pathway and dependence on the distance and
relative orientation of the molecules in question is presented. The
introduction of solvent effects is also discussed along with its implications
on the results, and the dissociation rate ($k_\mathrm{off}$) is derived from the
calculated energy barriers. Following this introduction, the selection of the
AB-AG system is described in detail. Next, a brief overview of the theoretical
methods adopted in this study is given, in particular the computational level,
the force field and the extent to which the solvent effects have been
introduced. In section \ref{ResultsDiscuss} the results are presented,
compared with the available experimental data, and discussed. The last section
is devoted to the conclusions.

\section{The Test Case}

\label{TestCase}

\begin{figure}
[h]
\includegraphics[width=0.8\columnwidth,clip]
{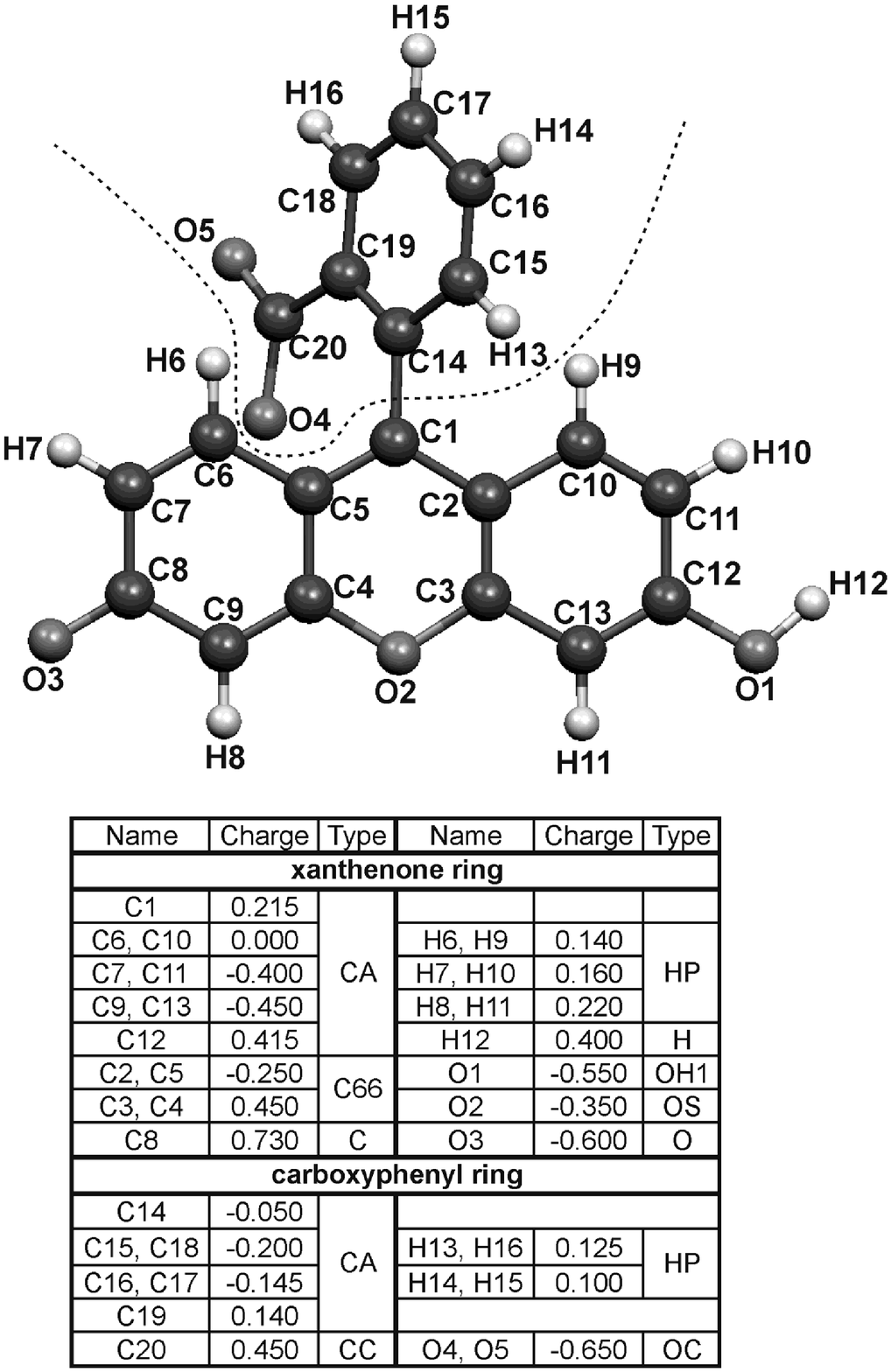}
\caption{Structural formula and assigned atom labels for Fluorescein \{2-(6-
hydroxy-3-oxo-(3H)-xanthen-9yl) benzoic acid\}. The force-field atom types
(see sub-section \ref{FluParm}) and the partial charges (units of \textit{e}) are listed in
the table underside. The dashed line puts in evidence the two aromatic (ring)
fragments labelled and grouped in the table.}
\label{F2}
\end{figure}

Fluorescein (Flu) is a synthetic hapten. It is extensively used in
fluorescence-based kinetic measurements of off-rates ($k_\mathrm{off}$)
\cite{VossA01}, and a valuable reference system for the understanding of
important immunological issues. Anti-fluorescein AB-AG complexes are also
clear-cut models in the sense that Flu is a small inert and rigid ligand (see
Figure \ref{F2}) and the off-rates of a number of anti-Flu complexes have been
found to display an Arrhenius-like behaviour \cite{SchwesingerA01}.%

The current study has been carried out for the anti-fluorescein IgG monoclonal
antibody \mbox{4-4-20} (\mbox{mAb4-4-20}), one of the most extensively studied by
thermodynamic, kinetic, structural, spectroscopic, and mutational methods
(\cite{MidelfortA01}, and references within), and for which two
crystallographic structures of its Fab region (highlighted in Figure \ref{F1})
have already been reported \cite{HerronA01,WhitlowA01}. A complete IgG~\mbox{mAb4-4-20}
molecule has two identical Fab fragments, each consisting of two constant
and two variable domains. The two variable domains (labelled V$_{L}$ and
V$_{H}$) constitute the so called Fv fragment (also highlighted in Figure
\ref{F1}), which is the minimal antigen-binding fragment. In fact, there are
many genetically engineered ABs that feature only the V$_{L}$ and V$_{H}$
domains \cite{JungA01}. This practice further endorses the idea of a system
with a restricted number of binding-determinant degrees of freedom. It also
makes it realistic (and computationally less demanding) to consider just the
mAb 4-4-20 variable domains: V$_{L}$ with 112 amino acids and V$_{H}$ with 118.

\section{Methods}
\label{Methods}

\subsection{Force field}
\label{FF}

Even reducing the system to the \mbox{mAb4-4-20} two variable domains plus Flu, it
amounts to \textit{ca.} 3600 atoms. It is, thus, too big to be computationally
addressed at any level of quantum mechanics. A realistic simplification is to
assume that the nuclei move in the average field created by all particles, and
use an empirical fit to this field -- an effective potential commonly known as
a force field. One then uses the computationally less demanding classical
mechanics formalism to calculate both static properties (equilibrium
structures, relative energies, etc.) and the time evolution of the system.

Much effort has been devoted to develop force fields suitable for studying
proteins, the CHARMM force field \cite{CHARMM22} (the one used in the present
work) being one of the most widely used nowadays. Its potential energy
function reads:%
\begin{align}
E&=\sum_{i=1}^{N_{r}}k_{i}^{r}(r_{i}-r_{i}^{0})^{2}+\sum_{i=1}^{N_{\theta
}}k_{i}^{\theta}(\theta_{i}-\theta_{i}^{0})^{2}\nonumber\\
& +\sum_{i=1}^{N_{\phi}}k_{i}^{\phi}\left[1+\cos(n_{i}\phi_{i}+\delta_{i})\right]\label{CHARMMff}\\
& +\sum_{i=1}^{N_{\chi}}k_{i}^{\chi}(\chi_{i}-\chi_{i}^{0})^{2}+%
\sum_{i=1}^{N_{S}}k_{i}^{S}(S_{i}-S_{i}^{0})^{2}\nonumber\\
& +\sum_{\genfrac{}{}{0pt}{}{{i,j=1}}{{i<j}}%
}^{N}\epsilon_{ij}\left[\left(\frac{R_{ij}}{r_{ij}}\right)^{12}-2\left(
\frac{R_{ij}}{r_{ij}}\right)^{6}\right]\nonumber\\
& +\sum_{\genfrac{}{}{0pt}{}{{i,j=1}}{{i<j}}%
}^{N}\frac{q_{i}q_{j}}{\varepsilon r_{ij}}.\nonumber
\end{align}
The energy $E$ is a function of the positions of all atoms in the system. The
first four summations are known as bonding terms: they extend to the
topologically defined $N_{r}$ covalent bonds `$r$', $N_{\theta}$ bond angles
`$\theta$', $N_{\phi}$ dihedral angles `$\phi$' and $N_{\chi}$ improper torsion
angles `$\chi$', respectively. For some specific bond angles, an additional
bonding term may be required as a function of the distance `$S$' between the
first and third atoms \cite{CHARMM22}. The term for the bonds describes the
energy required to deform a bond from its equilibrium value (denoted by the
subscript `$0$'), within a harmonic approximation, and an analogous
description holds for the remaining harmonic terms; the dihedral term is
chosen differently to satisfy the dihedral periodicity. The last two
summations of equation (\ref{CHARMMff}) are extended to all $N$ non-bonding
atom pairs $ij$ separated by three or more covalent bonds. The Lennard-Jones
6-12 potential term accounts for the van der Waals (vdW) interactions: for
each atom type (say $i$), there is a $R_{i}$ distance corresponding to a well
depth $\epsilon_{i}$, with $R_{i}=2^{1/6}\sigma_{i}$ and $\sigma_{i}$ the
distance for which the Lennard-Jones potential equals zero; the Lennard-Jones parameters
between pairs of different atoms are obtained from combination rules, the $\epsilon_{ij}$
values based on the geometric mean of $\epsilon_{i}$ and $\epsilon_{j}$ and $R_{ij}$
values form the arithmetic mean between $R_{i}$ and $R_{j}$. The Coulombic potential is
defined for the pairs of charges $q_{i}$ and $q_{j}$ separated by a distance
$r_{ij}$, and for a given dielectric constant $\varepsilon$ (the vacuum one by
default). The equilibrium values in the harmonic terms and the $R_{i}$ and
$\epsilon_{i}$ values are parameters derived from experimental data (\textit{e.g}.,
crystallographic structures) and \textit{ab initio} quantum mechanical calculations on
small reference molecules, presented and discussed in ref. \cite{CHARMM22}.
Partial charges ($q_{i}$, $q_{j}$) are also derived from such \textit{ab initio}
calculations.

\subsection{Implicit solvent}
\label{Implicit}

Proteins operate in aqueous solution. Solvation, stability and dissociation of
molecules in water are largely governed by electrostatic interactions. This is
particularly pertinent in proteins: more than 20\% of all amino acids in
globular proteins are ionized under physiological conditions and polar
side-chains occur in over another 25\% amino acids \cite{FogolariA01}.
However, introducing explicit water molecules in a computational simulation to
account for solvent effects dramatically increases the calculation time.
Moreover, when the calculations involve any energy minimisation-based
technique like calculating minimum energy reaction paths, the explicit water
molecules will arrange in a single conformation matrix, exerting forces on the
solute that are very different from the solvent mean force.

Alternatively, a continuum treatment of the solvent as a uniform dielectric
may provide an accurate enough description of such interactions, as long as
one accounts for the fact that a protein in aqueous solution (the
physiological medium) yields a system with two very different dielectric media
\cite{FogolariA02}. A most physically correct implicit solvent model arises
from solving the so-called Poisson--Boltzmann (PB) equation (see 
\cite{FogolariA02,BakerA01}, and references within). The protein
(macromolecule) is treated as a low-dielectric cavity bounded by the molecular
surface and containing partial atomic charges -- typically taken from the
classical molecular mechanics force field. The solvent (water) is implicitly
introduced by assuming a high-dielectric surrounding of the protein. And since
under physiological conditions macromolecules are dissolved in dilute saline
solutions (water with a dissolved electrolyte), a term for the average charge
density due to the mobile ions is also included. This classical continuum
electrostatics treatment relies on the (reasonable) assumption that it is
possible to replace the ionic potential of mean force with the mean
electrostatic potential, neglecting non-Coulombic interactions (\textit{e.g.}, vdW) and
ion correlations. The actual PB equation reads:%
\begin{align}
\nabla\cdot[\varepsilon(\mathbf{r})\nabla\varphi(\mathbf{r})]&=-4\pi\rho(\mathbf{r})\label{PBE}\\
& -4\pi\sum_{i=1}^{N}e\mathrm{q}_{i}n_{i}%
(\mathbf{r})\lambda(\mathbf{r}), \nonumber
\end{align}
with%
\begin{equation}
n_{i}(\mathbf{r})=n_{i}^{0}\exp(e\mathrm{q}_{i}\varphi(\mathbf{r})/k_\mathrm{B}\mathrm{T}).
\label{Boltzmann}%
\end{equation}
Equation (\ref{PBE}) relates the electrostatic potential $\varphi$
to the distribution of the protein atomic partial charges (charge density
$\rho$), the dielectric properties of both the protein and solvent
($\varepsilon$, position dependent $(\mathbf{r})$), and the charge density due
to the mobile ions given by the summation term; $q_{i}$ is the charge of ion
type $i$, $n_{i}(\mathbf{r})$\ its local concentration, $e$ the elementary
charge and $\lambda(\mathbf{r})$ a parameter that describes the ions'
accessibility at position $\mathbf{r}$. The boundary condition is
$\varphi(\mathbf{\infty})=0$. For each ion type, $n_{i}(\mathbf{r})$ is
described by a Boltzmann distribution (\ref{Boltzmann}) where $n_{i}^{0}$ is
the ion's concentration in bulk solution, $k_{B}$ the Boltzmann constant and
T the absolute temperature. As for the accessibility parameter, a general
consensus is that any point within one ionic radius from the macromolecule
is inaccessible (\textit{i.e.}, $\lambda(\mathbf{r})=0$), and it is implicit that the
region inside the macromolecular surface is inaccessible.
The remaining region outside has $\lambda (\mathbf{r})=1$. A typical value for the
ionic raidus is the one of Cl$^{-}$ (2~\AA), considering that
Na$^{+}$Cl$^{-}$ (sodium chloride) is a most frequently chosen electrolyte.

A description of the several possible approximations and numerical techniques
used to solve equation (\ref{PBE}) is beyond the scope of the present paper,
the reader being referred to the supporting literature of the software used in
this work, APBS (Adaptative Poisson--Boltzmann Solver)
\cite{BakerA01,SinghA01}. Briefly, the solute's charges are mapped onto a mesh
and the electrostatic potential in the presence of the dielectric continuum
solvent is determined at each point, via a finite difference numerical
solution of the PB equation. The mesh being a finite one, it is necessary to
set up the boundary potentials (at the lattice edge) accordingly. For the
present work, they are approximated by the sum of the Debye--H\"uckel potentials
of all the charges, meaning%
\begin{equation}
\varphi=\sum_{i=1}^{N}e\mathrm{q}_{i}\frac{\exp(-r_{i}/\lambda_{D}%
)}{\varepsilon_\mathrm{water}r_{i}}, \label{DH}%
\end{equation}
where $\lambda_{D}$, the Debye length, reads%
\begin{equation}
\lambda_{D}=\sqrt{\frac{\varepsilon_\mathrm{water}k_\mathrm{B}\mathrm{T}}{4\pi N_{A}\sum_{i=1}%
^{N}n_{i}^{0}e^{2}q_{i}^{2}}}. \label{DL}%
\end{equation}

\subsection{Fluorescein parameters}
\label{FluParm}

The available CHARMM parameterisation already has parameters for all amino
acids but not for fluorescein, so one has first to describe the later
consistently with the force field. In the present work, the required bonding
and Lennard-Jones parameters where derived by analogy to similar ones existing
in CHARMM. Partial atomic charges were fitted to reproduce the molecular
electrostatic potential (MEP) at selected points around the molecule according
to the Merz--Singh--Kollman scheme \cite{SinghA02,BeslerA01} implemented in the
Gaussian03 program \cite{Gaussian03}. The points are located in layers around
the molecule, the first layer corresponding to the van der Waals molecular
surface scaled by a factor of 1.4; the default scheme then adds three more
layers with scaling factors 1.6, 1.8 and 2.0. The MEP was generated at the
DFT/B3LYP level with the 6-31G(d) basis set. DFT (density functional theory)
is now a widely used and computationally convenient quantum mechanical method
well documented in many reference books (\textit{e.g.}, \cite{ParrB01}): it makes use
of exchange correlation functionals dependent on the electron density and its
gradient to tackle electron correlation effects. B3LYP was the functional of
choice and stands for the three-parameter Becke functional combined with the
Lee-Yang-Parr correlation functional \cite{ParrB01}.

For the quantum mechanical calculations, the coordinates for the starting Flu
conformation were taken from the complex crystal structure with the best
resolution (1.85 \AA \ \cite{WhitlowA01}), which has the coordinates deposited
in the RCSB Protein Data Bank \cite{BermanA01} with entry name 1FLR. Only the
acidic deprotonated form of Flu was considered, since this is the active form
in the experiments underlying the current study. The structure was energy
optimized before charge fitting. The charges were then further refined by
similarity to the set of already defined ones in the CHARMM force field (for
details on the approach see \cite{PaciA01,HenriquesA01,HenriquesA02}). The
ensuing Flu set of parameters was then used as an extension of the CHARMM
parameterisation, the corresponding CHARMM atom types and partial charges
being indicated in Figure \ref{F2}.

\subsection{Reference geometry}

\begin{figure}
[h]
\includegraphics[width=1.0\columnwidth,clip]
{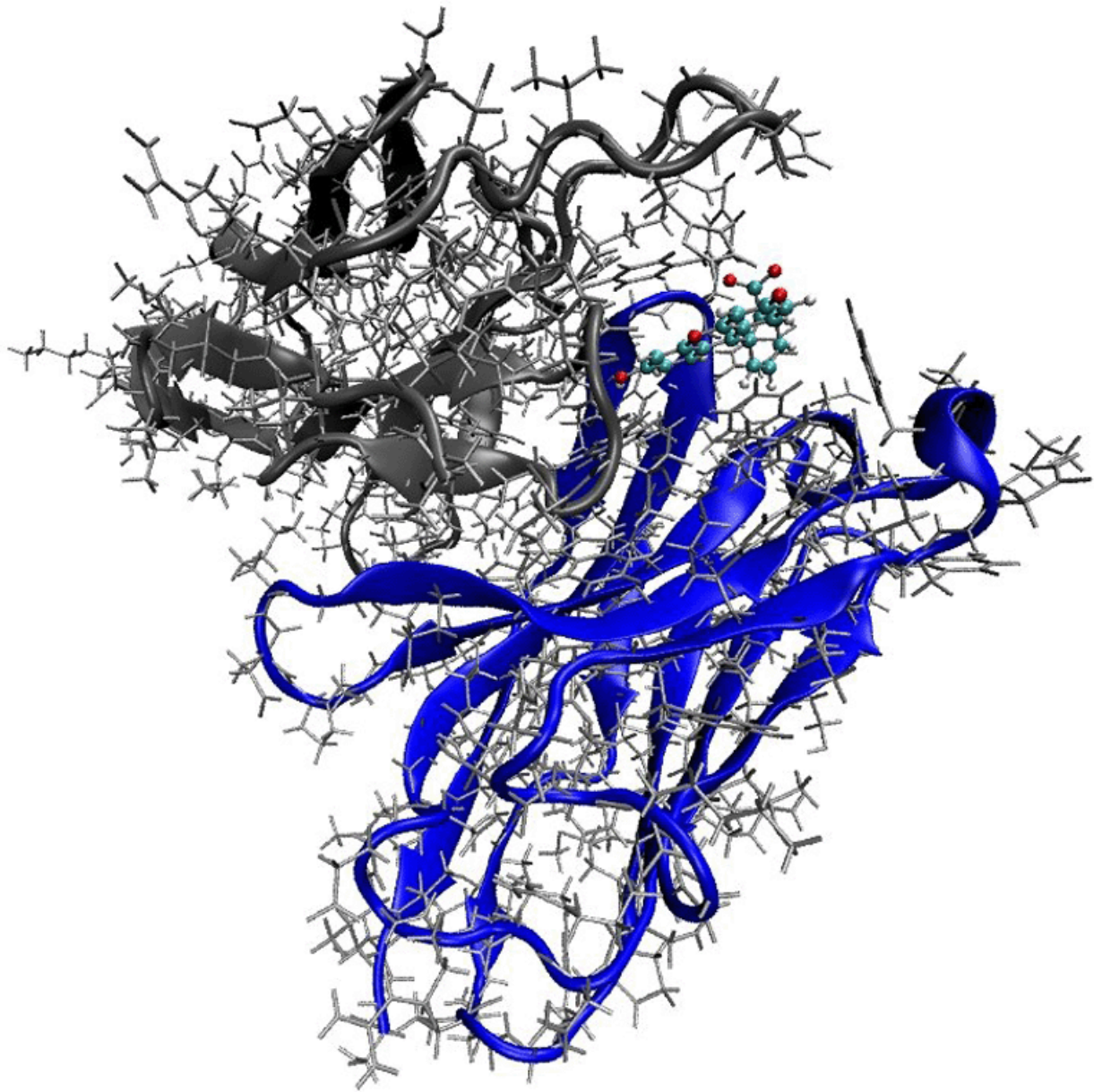}%
\caption{All-atom representation of the Fv-fragment of the \mbox{mAb4-4-20}-Flu
complex structure. The ribbon representation highlights the backbone of the
two chains, the H-chain in blue and the L-chain in gray. The Flu molecule is
depicted in ball-and-stick and coloured by element (CPK space-filling).}%
\label{F3}%
\end{figure}

The variable domains (V$_{L}$ and V$_{H}$) were extracted from the L and H
segments of the anti-Flu \mbox{4-4-20Fab} 1FLR crystal structure
\cite{WhitlowA01,BermanA01}. Crystallographic waters were stripped from the
structure and the C-terminal amino acids were capped with --NH$_{2}$
functional groups. A representation of the system is shown in Figure \ref{F3}. The
positions of the protein's missing hydrogens were initially guessed. Next, any
latent close contacts or anomalous bonding positions were cleared out by
relaxing the structure to an energy gradient tolerance of 0.05 eV$\cdot$\AA $^{-1}$,
at the classical mechanics level using the NAMD program
\cite{NAMD} with the extended CHARMM parameterisation. This relaxed structure
fully retains the experimental X-ray conformational features and it was used
as the starting conformation for the scanning. A full structure optimisation
(\textit{i.e.}, using a tighter energy gradient tolerance) was also carried
out but it introduced many small errors at the protein's secondary structure
level. This is because secondary structure relies on a network of backbone
hydrogen bonds, which are less accurately described in the framework of the
simplified molecular mechanics force field theories. The CHARMM energy
difference between the relaxed and fully minimised geometries is
$\sim$100 eV. The crystallographic structure itself has been resolved at
a temperature of $\sim$290 K \cite{WhitlowA01}, thus an estimate of the
corresponding average thermal energy (considering $k_{B}T$ per degree of
freedom) amounts to $\sim$270 eV for our simulation system. This indicates that the
full minimisation is only reaching some local minima. Considering the above
referred limitations of the force field, it is judicious to take the minimally
relaxed structure (closer to the X-ray one) as the reference structure for the
subsequent simulations.%

Flu is a particularly rigid molecule. Its essential degree of freedom is the
torsion around the bond between the xanthenone and the carboxyphenyl aromatic
rings (see Figure \ref{F2}), defining the angle between the planes of these
two rings. This angle has the value of $-$63$^{\circ}$
in both the crystal complexed form \cite{WhitlowA01} and the crystalline free
Flu \cite{DuBostA01}. For the above referred CHARMM energy relaxed structure,
the value of this angle is $-$67$^\circ$.
An energy optimization was also carried out for the hapten alone (without
the AB), the value for the angle in question being -62$^\circ$.
Moreover, the RMSD (root-mean-square deviation) between the crystallographic
and relaxed Flu bound structures is 0.201 \AA . These results are a good
indication of the validity of the derived set of CHARMM parameters.

\subsection{Distance scanning}
\label{SCAN}
In the pursuit for the suitable reaction coordinates to describe the system's
unbinding, the distance between the protein and Flu mass-centres could be a
first option, in close analogy to some cluster fission processes
\cite{ASolovyovIB01}. Yet for reasons that will become clear next, a distance
between two rationally selected atoms has been considered instead.

\begin{figure*}
[t]
\centering
\includegraphics[scale=0.80,clip]
{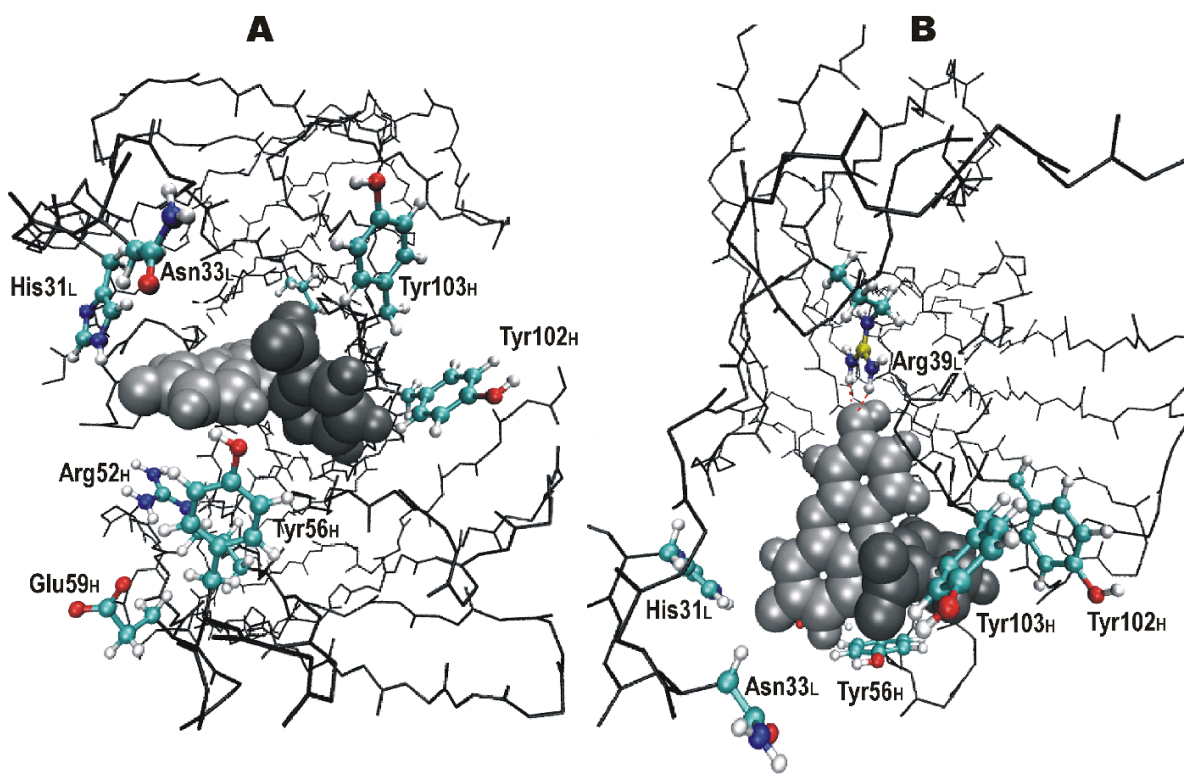}%
\caption{Fluorescein (van der Waals spheres in shades of grey) docked in the
antibody's binding cavity. The backbone of the antibody is depicted in black
sticks; the framing amino acid acids (labelled as described in the text) have
their side-chains displayed in coloured ball-and-stick. (\textbf{A}) front
view of the entrance of the cavity. (\textbf{B}) top view showing residue
Arg39$_{L}$ at the bottom of the cavity, with its atom C$_{\zeta}$ (zeta-carbon,
standard nomenclature) highlighted in yellow: the red dotted lines represent
hydrogen bonding between Arg39$_{L}$ and the hapten.}%
\label{F4}%
\end{figure*}

The shape of the binding pocket hosting Flu is most complementary to this
hapten, with a few amino acids at the rim of the pocket gating the entrance.
Superimposing the two available crystal structures results in an overall RMSD
of 0.419 \AA \ for the Flu atoms and 1.854 \AA \ for the protein, with a few
of those rim amino acids exhibiting some of the larger individual RMSD values
(up to 3.7 \AA ). Out of those, five amino acids -- His31$_{L}$, 
Asn33$_{L}$, Tyr56$_{H}$, Tyr102$_{H}$ and Tyr103$_{H}$ -- strategically
``frame'' amino acid Arg39$_{L}$ at the bottom of the pocket,
as depicted in Figure \ref{F4}. Amino acids are labelled%
(in the text and Figure \ref{F4}) according to the standard
amino acid 3-letter code (\cite{LehningerB01}; see also
{\tt http://www.chem.qmul.ac.uk/iupac/ AminoAcid/}), the number of the amino acid in
the 1FLR file and the chain identifier in subscript (\textit{e.g.}, Arg39$_{L}$, refers
to an Arginine that is numbered 39 in the L-chain). As highlighted in Figure
\ref{F4}-B, Arg39$_{L}$ has its +1-ionized group (centred on atom C$_{\zeta}$)
directly involved in hydrogen bonding to the hydroxyl group of Flu (atoms
O1--H12 in Figure \ref{F2}). Arg39$_{L}$ is also a mutation introduced during
the maturation process of \mbox{mAb4-4-20} \cite{JimenezA01}: the original residue
was a neutral, weakly polar Histidine. Upon this His-to-Arg mutation a slowing
in the unbinding of Flu by a 1.5-fold was experimentally observed
\cite{JimenezA01}, no doubt in consequence of the increased attraction between
the mutated amino acid 39$_{L}$ and Flu. Thus, it was only logical to consider
the distance between the groups of Arg39$_{L}$ and Flu engaged in that driving
hydrogen bonding as a most likely unbinding coordinate.

The distance between the C$_{\zeta}$ atom of Arg39$_{L}$ and the Flu's
hydroxyl oxygen (O1) was then set as the appropriate coordinate for scanning.
The scanning started from the distance in the reference conformation and
progressed in increments of 0.25 \AA \ until a $\sim$40 
\AA \ distance. At this distance and for the set cut-off, the interaction
energy between the hapten and the AB becomes zero. The scanning was also
performed for a few decreasing steps, \textit{i.e.}, for distances smaller
than the one in the reference structure. The distance value at each scanning
step was imposed by means of a strong harmonic constraint (force constant = 26
eV$\cdot$\AA $^{-2}$) between the referred oxygen and a dummy atom placed at the same
coordinates of the C$_{\zeta}$ atom. The need for a dummy atom arises from the
fact that, in CHARMM, the non-bonding energy of all atom pairs separated by
less than three covalent bonds is excluded \cite{CHARMM22}; the introduced
harmonic constraint is an ``artificial bond'' and therefore
should not be directly set between the C$_{\zeta}$ and OH atoms, otherwise
several pair interactions between Arg39$_{L}$ and Flu would be wrongly excluded.

For each scanning step, the system was energy minimised with NAMD to an energy
gradient tolerance $\leq$ 4$\times$10$^{-4}$ eV$\cdot$\AA $^{-1}$. A 12 \AA \ cut-off
on long-range interactions with a switch
smoothing function between 10 and 12 \AA \ was used. During minimisation, the
hapten was free to move (subject only to the scanning harmonic constraint)
while the AB was kept frozen for all but the side-chain atoms of a few key
amino acids gating the passage of the hapten. The unconstrained side-chains
belong to His31$_{L}$, Asn33$_{L}$, Arg52$_{H}$, Tyr56$_{H}$, Glu59$_{H}$,
Tyr102$_{H}$ and Tyr103$_{H}$. The reported energy of each minimised structure
was calculated after removing the referred harmonic constraint. In NAMD, it is
not possible to set two different dielectrics within the same system, so
minimisations were performed for $\varepsilon$ = 1, and solvent effects were
introduced as corrections \textit{a posteriori}, as described next.

For the final conformation of each scanning step, the electrostatic energy was
recalculated using the APBS program (refer to sub-section \ref{Implicit}).
The conformation of the last scanning step roughly
occupies a 70 \AA -side cubic box. The side was extended by an extra 20
\AA \ for solvent media, resulting in a 90$\times$90$\times$90 \AA $^{3}$ box
that was set equal for all scanning steps. Calculations were performed using
the APBS' adaptive refinement \cite{HolstA01}. A low dielectric constant of
$\epsilon$ = 2 was set for the macromolecule cavity
\cite{FogolariA02,PetersenA01} and the typical water value of $\epsilon
\approx80$ was set for the continuum solvent medium. The effect of a dilute
electrolyte in solution was assessed with a second run of calculations, for a
salt bulk concentration of 0.150 mol$\cdot$dm$^{-3}$ as in a typical physiological
media \cite{SinghA02}, and a temperature of 298 K.

\subsection{Exploring relative orientations}

\begin{figure}
[h]
\includegraphics[width=0.9\columnwidth,clip]
{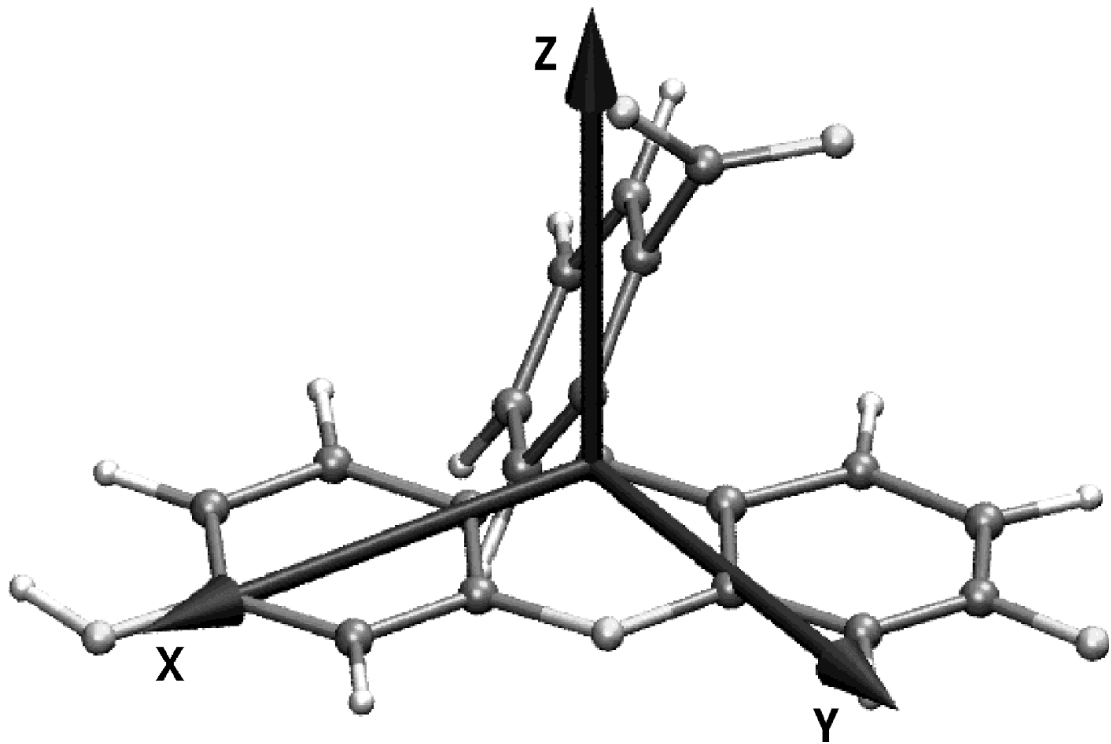}%
\caption{System of internal axes for Flu.}%
\label{F5}%
\end{figure}

The distance scanning scheme above described does not enforce an escaping
channel along a straight line, nor does it restrain the AB-hapten relative
orientations. For the sake of completion, a comprehensive overlook of the two
molecules relative position and mutual orientation should be performed. The
designated appropriate descriptors are the spherical coordinates ($r$,
$\theta$, $\phi$) and three Euler angles ($\alpha$, $\beta$, $\gamma$), for
which two coordinate frames are required. The referential frame, set as the
protein's principal axes of moment of inertia given that the protein is fixed
in space, and the moving-body local frame, \textit{i.e.} the Flu frame. Care
was taken to select this later, considering that during the scanning Flu does
not evolve in space as a completely rigid body. Its centre was set in Flu's
atom C1 since along the scanning the position of Flu's mass centre is
approximately coincident to this atom (0.2-0.4 \AA \ RMSD). The $xy$ plane was
made coincident to the rigid xanthenone ring, with the $x$ axis pointing in the
direction of atom O1, as displayed in Figure \ref{F5}.%

\subsection{Calculation of $k_\mathrm{off}$}

In compliance with the experimentally reported
Arrhenius-like behaviour \cite{SchwesingerA01} and within the context of the
reaction-rate theory \cite{HanggiA01}, the off-rate constant along the scanned
pathway was computed using the expression,
\begin{equation}
k_\mathrm{off}=\omega\exp(-\Delta E^{\ddag}/k_\mathrm{B}\mathrm{T}) \label{Koff}%
\end{equation}
where $\omega$ is the pre-exponential factor determining how frequently the
system approaches the barrier, and $\Delta E^{\ddag}$\ is the activation
energy (\textit{i.e.}, the barrier height). The harmonic approximation was used to
estimate the Arrhenius-like pre-factor. It reads%
\begin{equation}
\omega=\frac{1}{2\pi}\sqrt{k/\mu} \label{Kforce},%
\end{equation}
where $\mu$ is the reduced mass of the system and $k$ the harmonic force constant.
This latter was obtained from parabolic fit of the data (the bounding region of the well in
the energy profiles) using the Mathematica\textregistered\ software package. For systems
similar to the one presented here (with reduced masses in the 200-500 range and binding
pocket's length within 3-7 \AA), an estimation of the frequency $\omega$ falls in the
$10^{11}$-$10^{12}$ s$^{-1}$ range. 
%\textcolor{red}{\ldots}

\section{Results and Discussion}

\label{ResultsDiscuss}

\subsection{Energetic and structural analysis}

\begin{figure}
[h]
\includegraphics[width=0.9\columnwidth]
{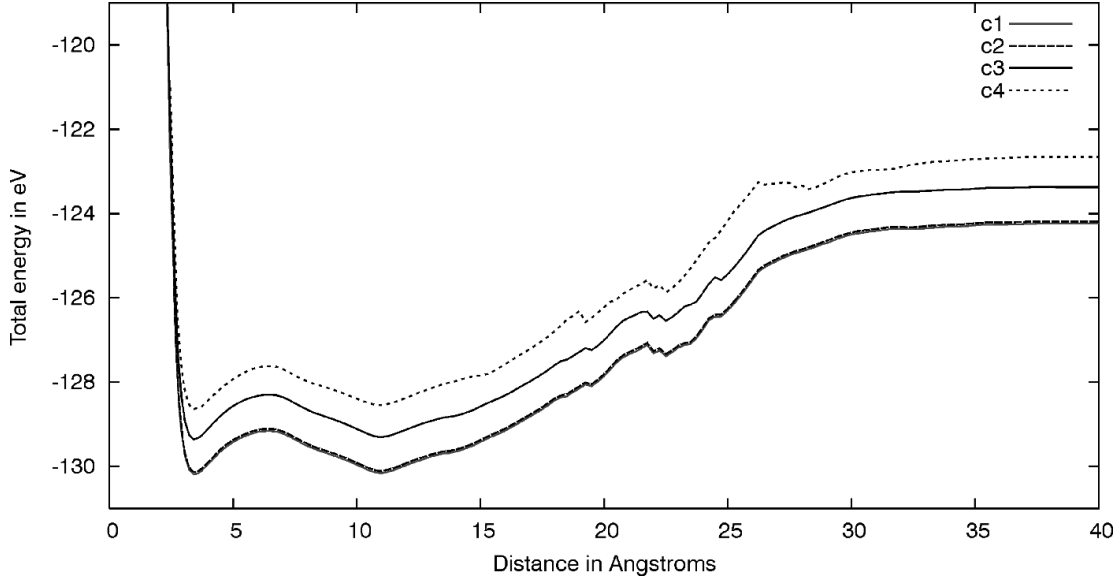}
\caption{Energy profiles (\textit{in vacuo}) for different distance scanning
runs, corresponding to different constraining schemes on the protein atoms.
Each curve corresponds to a different number of gating amino acid side-chains
that were allowed to move during each scanning, namely seven side chains (c1),
six (c2) five (c3) and four (c4) (details in the text).}
\label{F6}
\end{figure}

\begin{figure}
[h]
\includegraphics[width=0.9\columnwidth]
{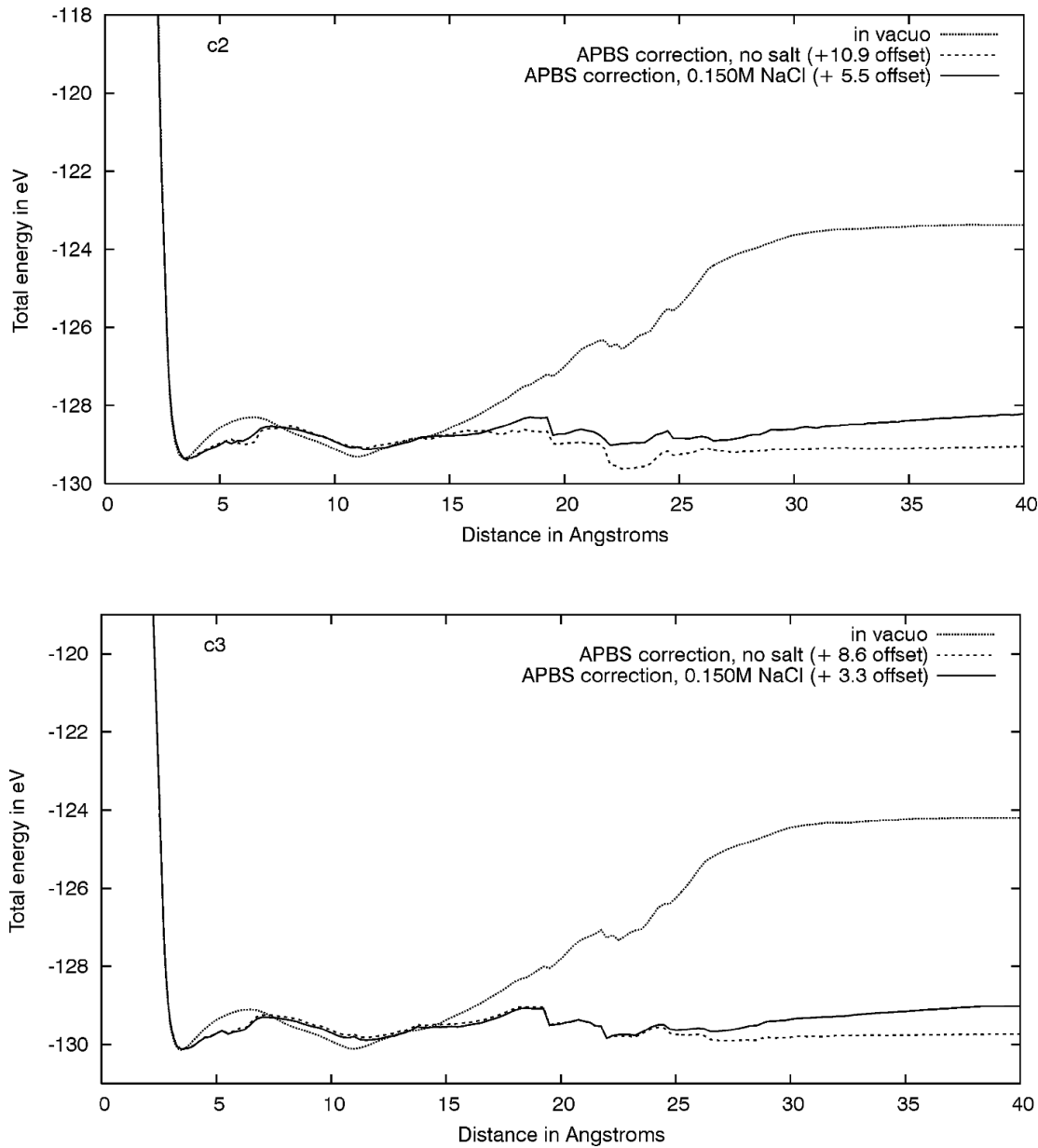}
\caption{Comparison of the distance scanning energy profiles \textit{in
vacuo} and with implicit solvent corrections (with and without dissolved
electrolyte), for the `c2' and `c3' constraining schemes.}
\label{F7}
\end{figure}

The energy profiles resulting from the scanning runs with and without solvent
correction are plotted in Figures \ref{F6} and \ref{F7}.%

Figure \ref{F6} displays the \textit{in vacuo} results for four different scanning runs,
corresponding to different constraining schemes on the protein atoms. The scheme
referring to the seven unconstrained side-chains enumerated in sub-section \ref{SCAN}
has been labelled `c1' in Figure \ref{F6}. To better assess on the influence of those
7 amino acids on the escaping profile, they were subject to successive
constraining procedures, exemplified in Figure \ref{F6} for three representative
cases, labelled `c2', `c3' and `c4', that correspond to six, five and four
unconstrained side chains (out of the initial seven). The plotted `c3' curve,
for instance, results from moving only the side-chains of the 5 gating residues
sginaled out in the second paragraph of sub-section \ref{SCAN} and visible in Figure \ref{F4}-B.

The constraining limit is the set of amino acids Asn33$_{L}$, Tyr56$_{H}$,
Tyr102$_{H}$ and Tyr103$_{H}$ corresponding to the `c4' curve. Within this
limit, no general significant differences on the energy profile arise from the
explored different schemes. These 4 amino acids always experience significant
conformational changes upon the hapten's passage, in comparison to the
remaining moving ones which just slightly adjust positioning. The plane
defined by the side-chain oxygens of the 4 amino acids in question can be
taken as the outmost limit of the protein's pocket, and it is intersected at a
$\sim$15 \AA \ scanning distance. Below this separation distance, the
total energy plots in Figure \ref{F6} depict the expected profile for an activated
process. For the different curves, the height and shape of the energetic
barrier at $\sim$7 \AA \ is essentially the same: 1.029, 1.027, 1.060 and 1.026 eV
respectively for 7, 6, 5 and 4 moving side-chains. Past the 15 \AA \ distance,
the \textit{in vacuo} profiles depict an asymptotic increase to a final
plateau above the referred energetic barrier, making unbinding unfeasible.
Predictably, the inclusion of solvent effects rectify the asymptotic behaviour
depicted in the \textit{in vacuo} profiles, as exemplified in Figure \ref{F7} for two scanning runs.
At larger separation distances the energy profile has been significantly
flattened, ant it is also for the larger distances that the effect of the
dissolved electrolyte becomes perceptible. In solution, the electrostatic
interactions between the protein and the escaping hapten are effectively
screened by the high-dielectric, allowing for unbinding to happen. Of
relevance is also the decrease in the height of the energetic barrier at
\ $\sim$7 \AA : with implicit solvent effects, this barrier value is
0.863 and 0.871 eV, respectively for the `c2' and `c3' schemes.

\begin{figure*}[t]
\centering
\includegraphics[scale=0.8,clip]
{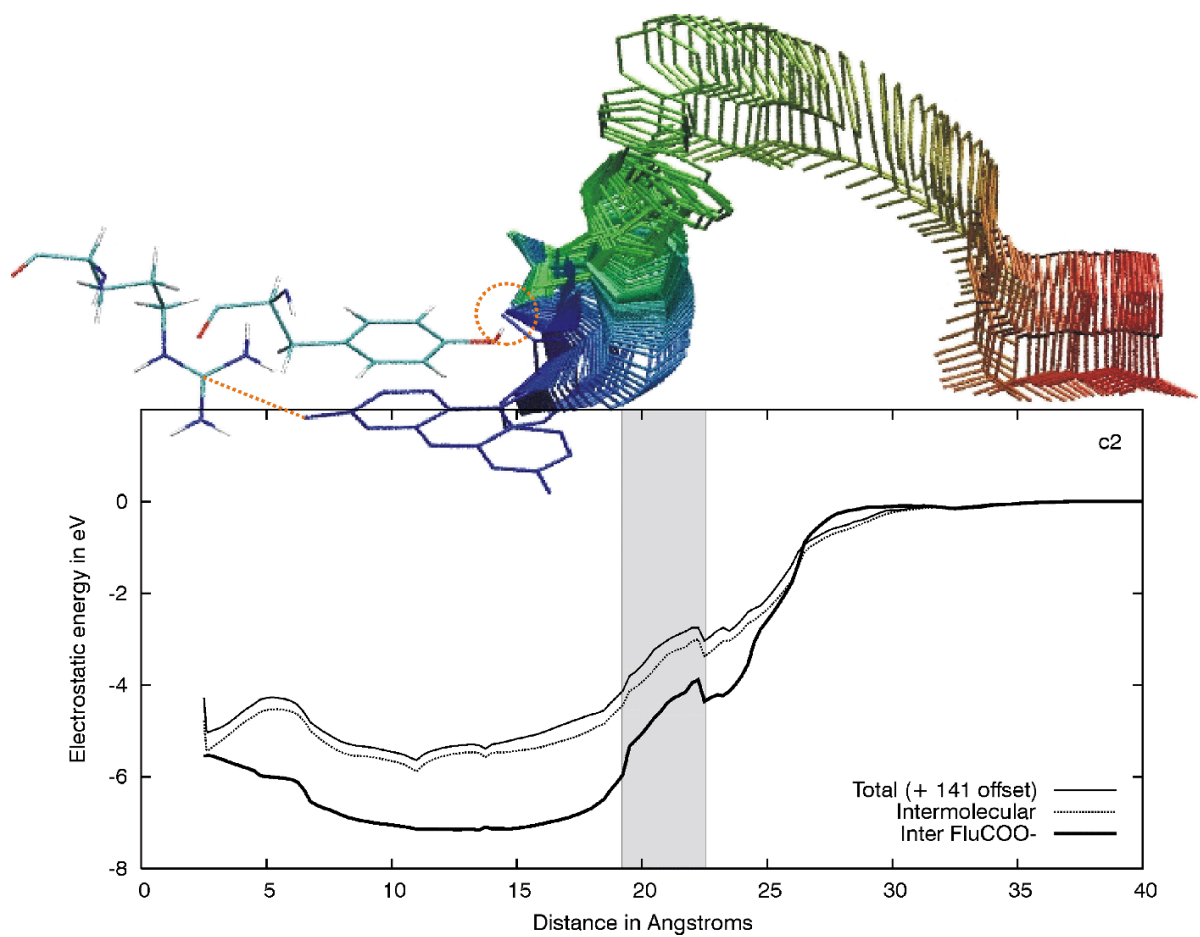}
\caption{Electrostatic energy profiles (\textit{in vacuo}) for the overall
system, its intermolecular component and the interaction energy between the
protein and the COO$^{-}$ group (atoms C20, O4, O5 of the Flu's carboxyphenyl ring;
see Figure \ref{F2}). The picture on top portraits the path of that ring along
the scanning: the dashed yellow straight line puts in evidence the coordinate
being scanned while the circle emphasises the anchor point of the COO$^{-}$
group at the protein' surface (see details in the text). The grey area puts in
evidence a jagged region in the profile.}
\label{F8}
\end{figure*}

\begin{figure}
[h]
\includegraphics[width=0.8\columnwidth]
{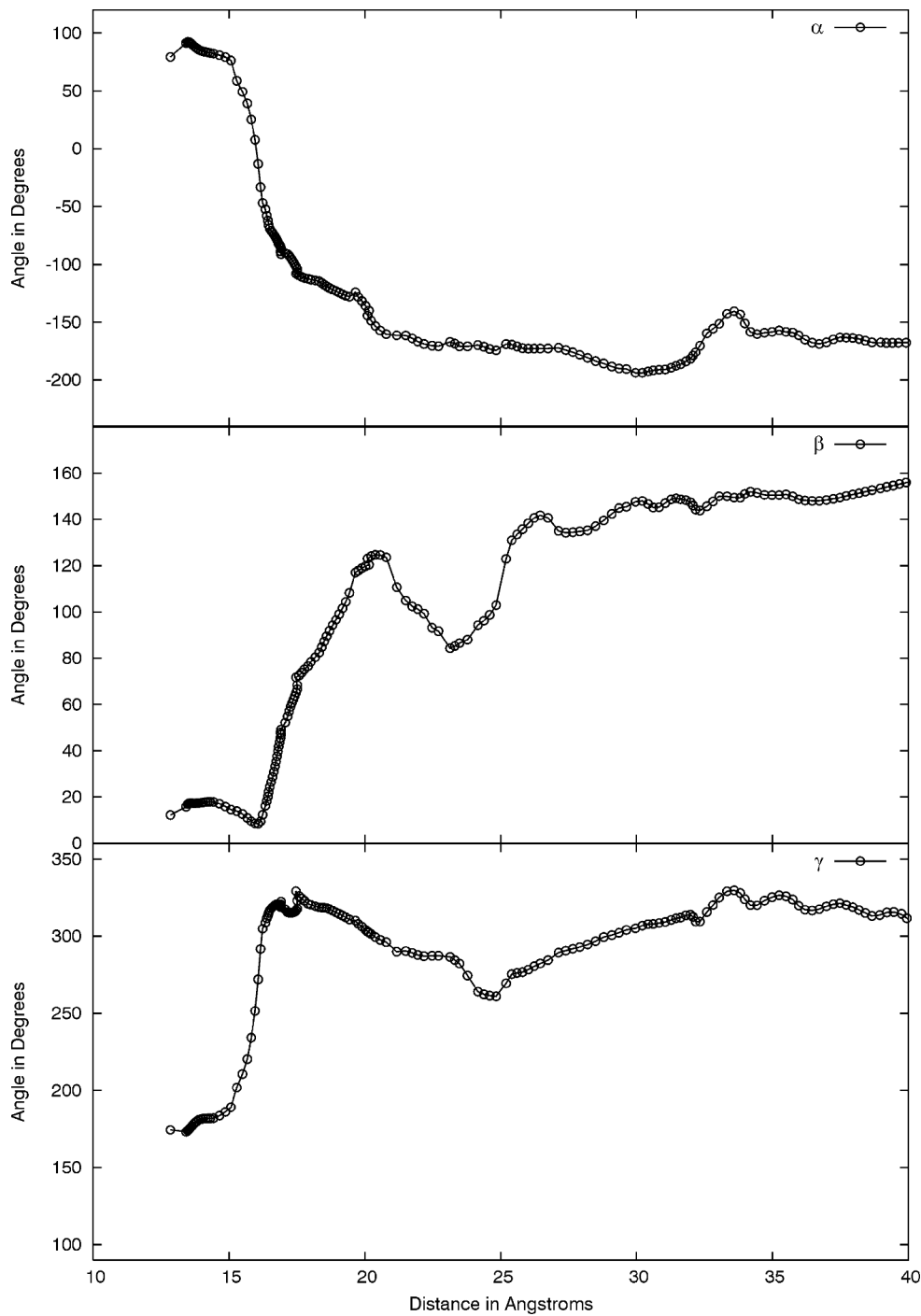}
\caption{Euler angles as a function of the scanned distance coordinate.}
\label{F9}
\end{figure}

The jagged contour emerging at $\sim$20 \AA \ also deserves some
attention. A detailed analysis of the energetic components was carried out,
with particular emphasis on the Coulombic component, as exemplified in Figure
\ref{F8} for scanning run `c2'.%
A previous work had already shown that, at larger distances, electrostatics play a
major role in fragmentation \cite{ISolovyovA01}. By comparing Figures \ref{F7} and \ref{F8},
one perceives the jagged pattern similarities between the total energy and its electrostatic component.
Moreover, the electrostatic energy and its intermolecular component --
\textit{i.e.}, the electrostatic interaction energy between the protein and
the hapten -- run parallel. To this intermolecular energy, a major
contribution comes from the COO$^{-}$ group of Flu. This group gets anchored
via hydrogen-bonds at the protein's surface as the hapten leaves the pocket:
the anchor point is yellow circled in the top picture of Figure \ref{F8}. The hapten
rotates around this point as the scanning distance is further increased, until
it finally detaches from the surface. The detachment features a somewhat
irregular trajectory of the escaping hapten: it is the region of the top
picture in Figure \ref{F8}, right above the grey area highlighting the jagged contour
in the plot. A better perception of the hapten's rotation as it leaves the
binding pocket can be gained from the plotting of the Euler angles along the
scanning, presented in Figure \ref{F9}. The steeper variation of the angles in the
15-20 \AA \ region corresponds to the anchoring track of the COO$^{-}$ group
at the protein' surface. As the hapten detaches, a swift change in its
orientation is observed, made evident by the plots for the Euler angles from
$\sim$20 \AA \ on. At this stage, one can not ascertain whether or not
this pronounced ``trapping'' of the hapten to the protein'
surface is a genuine feature of the unbinding. That would at least require the
other known mutations of the anti-fluorescein mAb4-4-20 to be subject to an
analogous study, which is beyond the scope of the present paper.%

\begin{figure}
[h]
\includegraphics[width=0.9\columnwidth]
{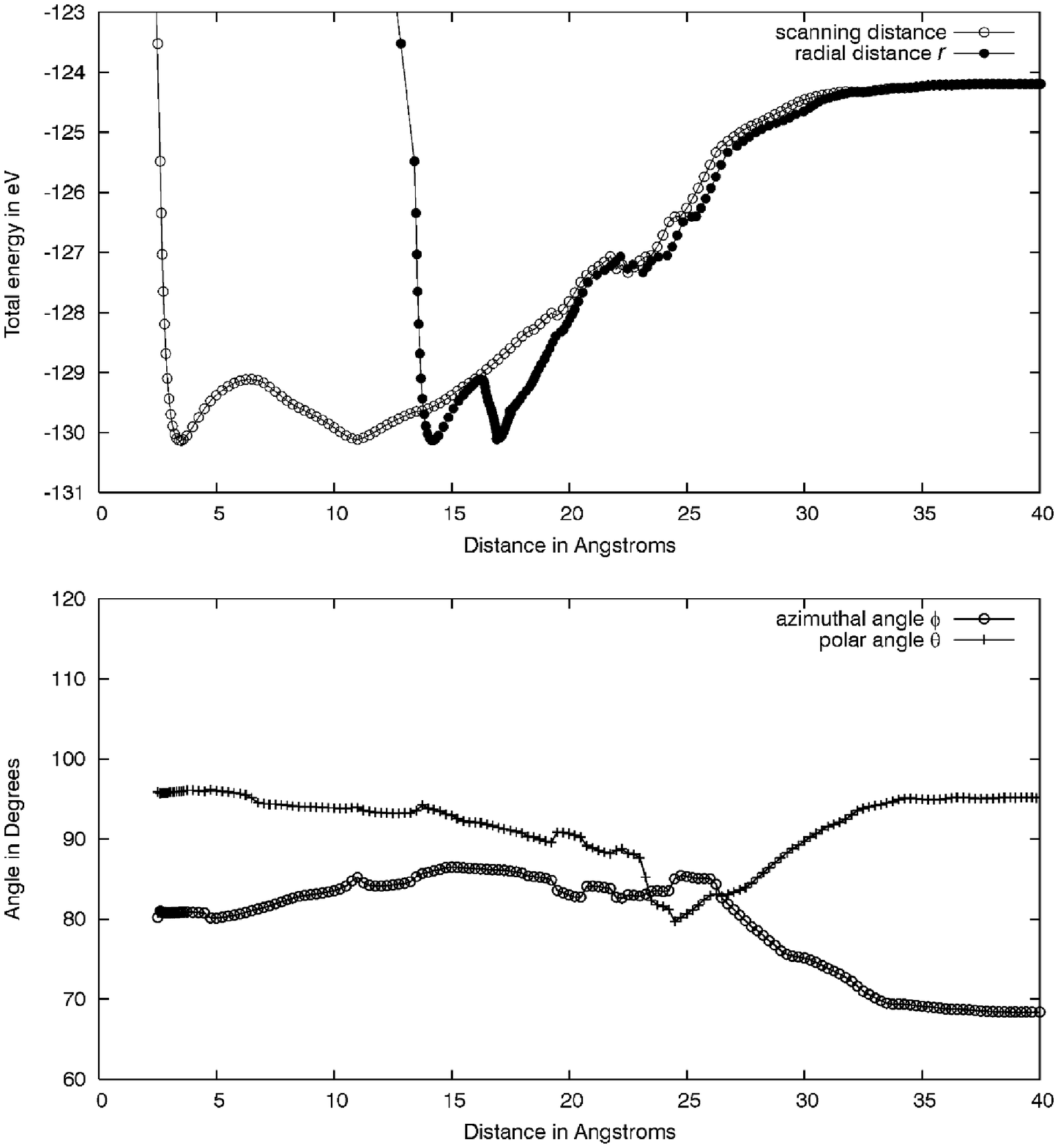}
\caption{Spherical coordinates along the `c2' scanning run. The total energy
of the system is plotted as a function of both the radial distance ($r$) and the
scanned distance coordinate. The angles $\theta$ and $\phi$ are plotted only
as a function of the scanning distance coordinate.}
\label{F10}
\end{figure}

The fact remains that, on the overall, the total energy profiles are smooth
(without discontinuities), as clear from the top plot in Figure \ref{F10} of the
energy as a function of both the scanning coordinate and the radial distance
($r$). They portray a plausible unbinding channel, provided solvent effects
are included, though one can not claim that they correspond to the minimum
energy pathway. One possible step to assess that would be to perform a more
comprehensive probing of the positional/orientational space of the hapten --
beyond the points determined by the presently selected reaction coordinate.
Yet, such a study involves a substantial computational effort, even if
confined to some plausible escaping window in space. On the other hand, the
presently computed profiles can be used to derive $k_\mathrm{off}$, and by comparison
to the corresponding experimental values(s), a first evaluation of the
scanning approach here introduced can be made.
 
\subsection{$k_\mathrm{off}$ determination}

Table I presents the calculated values of $k_\mathrm{off}$, with and without solvent
correction, resulting from parabolic fit to the profiles (0.99 $\leq$ R$^{2}$
$\geq$ 0.87), considering the energy barrier at $\sim$7
\AA \ (\textit{vide supra}). Experimentally available $k_\mathrm{off}$ values are also
presented for comparison. It becomes immediately evident that, even for an
extensively studied system like the \mbox{mAb4-4-20}--fluorescein one, experimental
$k_\mathrm{off}$ values may differ by an order of magnitude, depending on setup
conditions and techniques \cite{JimenezA01,MummertA01,BoderA01}.
As for our estimated values, while the \textit{in vacuo}
results are off-range, the solvent-corrected ones are comparable to the
experimental results. The equilibrium distance between the antibody and the
hapten (the well minimum) also compares better to the experimental value in
the case of the solvent-corrected simulations. Finally, remark that the
different constraining schemes have little influence on the order of magnitude
of the $k_\mathrm{off}$ values.

\begingroup
\begin{table*}[t]
\caption{Kinetic and equilibrium parameters obtained from calculations based
on the computational scanning. Available experimental values are also
presented for comparison: (a) and (c) determined in solution (ref.
\cite{MummertA01} and \cite{BoderA01} respectively), and (b)
at a surface by SPR \cite{JimenezA01}.} \label{T1}
\begin{ruledtabular}
\begin{tabular}[c]{c|c|c|c|c|c|c}
\multirow{2}{*}{parameter} & \multicolumn{4}{c|}{simulations} & \multirow{2}{*}{experimental}
& \multirow{2}{*}{T(K)}\\
\cline{2-5}
& \multicolumn{2}{c|}{\textit{in vacuo}} & \multicolumn{2}{c|}{solvent corrected} & & \\
\hline
& c2 & c3 & c2 & c3 &  & \\
\cline{2-5}
&  &  &  &  & $1.9\times10^{-3}$ $^{(a)}$ & 291\\
$k_\mathrm{off}$(s$^{-1}$) & $3.4\times10^{-6}$ & $6.8\times10^{-6}$ &
$4.1\times10^{-3}$ & $5.4\times10^{-3}$ & $6.8\times10^{-3}$ $^{(b)}$ & 298\\
&  &  &  &  & $4.3\times10^{-3}$ -- $2.5\times10^{-2}$ $^{(c)}$ & 298\\
\hline
equilibrium & \multirow{2}{*}{3.50} & \multirow{2}{*}{3.55}
& \multirow{2}{*}{3.60} & \multirow{2}{*}{3.65} & \multirow{2}{*}{3.65} & \multirow{2}{*}{291}\\
distance (\AA ) & & & & & & \\
\end{tabular}
\end{ruledtabular}
\end{table*}\endgroup

\section{Concluding Remarks}

\label{Conclusion}

Here we presented our first attempt to describe the unbinding of a complex
biomolecular system in terms of a reduced set of relevant generalized
coordinates while restricting most of its conformational internal degrees of
freedom. The reported results open a practical and physically sound procedure
to compute energy profiles along the selected reaction coordinate(s). This was
demonstrated in the present work for an experimentally well studied complex of
the biologically relevant antigen-antibody system. For the example in
question, it was actually possible to find a distance dependent escaping
channel in the multidimensional potential energy landscape, thus reducing the
unbinding to a low-dimensional problem: the system seems to be efficiently
bound by this one distance coordinate. The effect of the solvent was also
accounted for, and despite the fact that it was introduced as a correction
\textit{a posteriori}, it allowed us to ascertain that this is one effect that
needs to be included, for it has a significant influence in the overall
energetic profile and subsequent parameters derived from it. With solvent
effects, the derived off-rates are in reasonable agreement with the
experimentally determined ones, a result that can be regarded as an indicator
that ours is indeed a realistic approach.

The proposed approach would no doubt benefit from further refinements, namely
in the way solvent effects are introduced (\textit{viz.} include them during scanning,
both implicit and explicitly) and in the kinetic model, which we intend to carry on
in the near future. We also have in mind to apply this same approach to the maturation
series and engineered mutants of the \mbox{4-4-20}--fluorescein complex. That would
allow us to further test our strategy,
in particular its sensitiveness to the energetic differences arising from
antibody's single-point mutations. In addition, computed association rates
would be valuable parameters in different areas of immunological research,
namely theoretical immunology \cite{HermannA01}. Remark also that calculated
$k_\mathrm{off}$ values could be used to determine the related association rate
$k_\mathrm{on}$  using the relation $k_\mathrm{on}=$ $k_\mathrm{off}/K_\mathrm{d}$%
\cite{JimenezA01}, for those systems where only the equilibrium dissociation
constant $K_{d}$ has been experimentally measured. And by identifying and rationalize
the involved key structural features and interactions determining the unbinding, one
could make insightful predictions and propose, for instance, affinity-enhancing
mutations. Our long term goal is to extend our research to other molecular
recognition processes besides the antigen-antibody one and to test the
applicability and universality of our approach.
\vspace{0.1 cm}
\begin{acknowledgments}
Financial support from the NoE EXCELL EU project is gratefully acknowledged.
The authors also thank Ilia A. Solov'yov for his valuable assistance in the
implementation of the Euler angle's analysis, and Dr. Michael Meyer-Hermann
for drawing our attention to the importance of the problem considered for 
theoretical immunology. Finally, we are grateful to Prof. Walter Greiner for useful
discussions.
\end{acknowledgments}

\bibliographystyle{unsrt}
\bibliography{eshenriques_references}

\begin{thebibliography}{10}

\bibitem{ASolovyovIB01}
A.~V. Solov'yov, J.-P. Connerade, and W.~Greiner.
\newblock {\em Latest Advances in Atomic Cluster Collisions}.
\newblock Imperial College Press, London, 2004.

\bibitem{LyalinA01}
A.~Lyalin, O.~I. Obolensky, A.~V. Solov'yov, and W.~Greiner.
\newblock {\em Int. J. Modern Phys. E}, 15:153--195, 2006.

\bibitem{ISolovyovA01}
I.~A. Solov'yov, A.~V. Yakubovich, A.~V. Solov'yov, and W.~Greiner.
\newblock {\em J. Exp. Theo. Phys.}, 103:463--471, 2006.

\bibitem{WedemayerA01}
G.~J. Wedemayer, P.~A. Patten, L.~H. Wang, P.~G. Schultz, and R.~C. Stevens.
\newblock {\em Science}, 276:1665--1669, 1997.

\bibitem{LehningerB01}
David~L. Nelson and Michael~M. Cox.
\newblock {\em Lehninger Principles of Biochemistry}.
\newblock W.H. Freeman and Company, New York, 2005.

\bibitem{ManserA01}
Tim Manser.
\newblock Textbook germinal centers?
\newblock {\em The Journal of Immunology}, 172:3369--3375, 2004.

\bibitem{SchwesingerA01}
F.~Schwesinger, R.~Ros, T.~Strunz, D.~Anselmetti, H.-J. Güntherodt,
  A.~Honegger, L.~Jermutus, L.~Tiefenauer, and A.~Pl{\"u}ckthun.
\newblock {\em Proc. Natl. Acad. Sci. U.S.A.}, 97:9967--9971, 2000.

\bibitem{FooteA01}
J.~Foote and H.~N. Eisen.
\newblock {\em Proc. Natl. Acad. Sci. U.S.A.}, 92:1254--1256, 1995.

\bibitem{JimenezA01}
R.~Jimenez, G.~Salazar, T.~Joo J.~Yin, and F.~E. Romesberg.
\newblock {\em Proc. Natl. Acad. Sci. U.S.A.}, 101:3803--3808, 2004.

\bibitem{HinterdorferA01}
P.~Hinterdorfer, W.~Baumgartner, H.~J. Gruber, K.~Schlicher, and H.~Schindler.
\newblock {\em Proc. Natl. Acad. Sci. U.S.A.}, 93:3477--3481, 1996.

\bibitem{DammerA01}
U.~Dammer, M.~Hegner, D.~Anselmetti, P.~Wagner, M.~Dreier, W.~Huber, and H.-J.
  G{\"u}ntherodt.
\newblock {\em Biophys. J.}, 70:2437--2441, 1996.

\bibitem{AllenA01}
S.~Allen, X.~Chen, J.~Davies, M.~Davies, A.~C. Dawkes, J.~C. Edwards, C.~J.
  Roberts, J.~Sefton, S.~J.~B. Tendler, and P.~M. Williams.
\newblock {\em Biochemistry}, 36:7457--7463, 1997.

\bibitem{SulchekA01}
T.~A. Sulchek, R.~W. Friddle, E.~Y.~Lau K.~Langry, H.~Albrecht, T.~V. Ratto,
  S.~J. DeNardo, M.~E. Colvin, and A.~Noy.
\newblock {\em Proc. Natl. Acad. Sci. U.S.A.}, 102:16638--16643, 2005.

\bibitem{GrubmullerIB01}
H.~Grubm{\"u}ller.
\newblock {\em Protein-Ligand Interactions}, chapter Force probe molecular
  dynamics simulations, pages 493--515.
\newblock The Human Press Inc., NJ USA, 2005.

\bibitem{VossA01}
E.~W.~Voss Jr.
\newblock {\em J. Mol. Recognit.}, 6:51--58, 1993.

\bibitem{MidelfortA01}
K.~S. Midelfort, H.~H. Hernandez, S.~M. Lippow, B.~Tidor, C.~L. Drennan, and
  K.~D. Wittrup.
\newblock {\em J. Mol. Biol.}, 343:685--701, 2004.

\bibitem{HerronA01}
J.~N. Herron, X.~M. He, M.~L. Mason, E~W.~Voss Jr., and A.~B. Edmundson.
\newblock {\em Proteins}, 5:271--280, 1989.

\bibitem{WhitlowA01}
M.~Whitlow, A.~J. Howard, J.~F. Wood, E.~W.~Voss Jr., and K.~D. Hardman.
\newblock {\em Protein Eng.}, 8:749--761, 1995.

\bibitem{JungA01}
S.~Jung and A.~Pluckthun.
\newblock {\em Protein Eng.}, 10:959--966, 1997.

\bibitem{CHARMM22}
A.~D.~MacKerell Jr., D.~Bashford, M.~Bellott, R.~L.~Dunbrack Jr., J.~Evanseck,
  M.~J. Field, S.~Fischer, J.~Gao, H.~Guo, S.~Ha, D.~Joseph, L.~Kuchnir,
  K.~Kuczera, F.~T.~K. Lau, C.~Mattos, S.~Michnick, T.~Ngo, D.~T. Nguyen,
  B.~Prodhom, W.~E.~Reiher III, B.~Roux, M.~Schlenkrich, J.~Smith, R.~Stote,
  J.~Straub, M.~Watanabe, J.~Wiorkiewicz-Kuczera, D.~Yin, and M.~Karplus.
\newblock {\em J. Phys. Chem. B}, 102:3586--3616, 1998.

\bibitem{FogolariA01}
F.~Fogolari, A.~Brigo, and H.~Molinari.
\newblock {\em J. Mol. Recognit.}, 15:377--392, 2002.

\bibitem{FogolariA02}
F.~Fogolari, P.~Zuccato, G.~Esposito, and P.~Viglino.
\newblock {\em Biophys. J.}, 76:1--16, 1999.

\bibitem{BakerA01}
N.~A. Baker, D.~Sept, Joseph S., M.~J. Holst, and J.~A. MacCammon.
\newblock {\em Proc. Natl. Acad. Sci. U.S.A.}, 98:10037--10041, 2001.

\bibitem{SinghA01}
U.~C. Singh and P.~A. Kollman.
\newblock {\em J. Comput. Chem.}, 5:129--145, 1984.

\bibitem{SinghA02}
B.~P. Singh, H.~B. Bohidar, and S.~Chopra.
\newblock {\em Biopolymers}, 31:1387--1396, 1991.

\bibitem{BeslerA01}
B.~H. Besler, K.~M.~Merz Jr., and P.~A. Kollman.
\newblock {\em J. Comput. Chem.}, 11:431--439, 1990.

\bibitem{Gaussian03}
Gaussian 03:~Revision C.02, M.~J. Frisch, G.~W. Trucks, H.~B. Schlegel, G.~E.
  Scuseria, M.~A. Robb, J.~R. Cheeseman, J.~A.~Montgomery Jr., T.~Vreven, K.~N.
  Kudin, J.~C. Burant, J.~M. Millam, S.~S. Iyengar, J.~Tomasi, V.~Barone,
  B.~Mennucci, M.~Cossi, G.~Scalmani, N.~Rega, G.~A. Petersson, H.~Nakatsuji,
  M.~Hada, M.~Ehara, K.~Toyota, R.~Fukuda, J.~Hasegawa, M.~Ishida, T.~Nakajima,
  Y.~Honda, O.~Kitao, H.~Nakai, M.~Klene, X.~Li, J.~E. Knox, H.~P. Hratchian,
  J.~B. Cross, C.~Adamo, J.~Jaramillo, R.~Gomperts, R.~E. Stratmann, O.~Yazyev,
  A.~J. Austin, R.~Cammi, C.~Pomelli, J.~W. Ochterski, P.~Y. Ayala,
  K.~Morokuma, G.~A. Voth, P.~Salvador, J.~J. Dannenberg, V.~G. Zakrzewski,
  S.~Dapprich, A.~D. Daniels, M.~C. Strain, O.~Farkas, D.~K. Malick, A.~D.
  Rabuck, K.~Raghavachari, J.~B. Foresman, J.~V. Ortiz, Q.~Cui, A.~G. Baboul,
  S.~Clifford, J.~Cioslowski, B.~B. Stefanov, G.~Liu, A.~Liashenko, P.~Piskorz,
  I.~Komaromi, R.~L. Martin, D.~J. Fox, T.~Keith, M.~A. Al-Laham, C.~Y. Peng,
  A.~Nanayakkara, M.~Challacombe, P.~M.~W. Gill, B.~Johnson, W.~Chen, M.~W.
  Wong, C.~Gonzalez, and J.~A. Pople.
\newblock {\em Gaussian Inc., Wallingford CT}, 2004.

\bibitem{ParrB01}
R.~G. Parr and W.~Yang.
\newblock {\em Density-Functional Theory of Atoms and Molecules}.
\newblock Oxford University Press, New York, 1994.

\bibitem{BermanA01}
H.~Berman, K.~Henrick, H.~Nakamura, and J.~L. Markley.
\newblock {\em Nucleic Acids Research: Database issue D1-D3
  doi:10.1093/nar/gkl971}, 00, 2006.

\bibitem{PaciA01}
E.~Paci, A.~Caflisch, A.~Pl{\"u}ckthun, and M.~Karplus.
\newblock {\em J. Mol. Biol.}, 314:589--605, 2001.

\bibitem{HenriquesA01}
E.~S. Henriques, M.~Bastos, C.~F.~G.~C. Geraldes, and M.~J. Ramos.
\newblock {\em Int. J. Quant. Chem.}, 73:237--248, 1999.

\bibitem{HenriquesA02}
E.~S. Henriques, M.~A.~C. Nascimento, and M.~J. Ramos.
\newblock {\em Int. J. Quant. Chem.}, 106:2107--2121, 2006.

\bibitem{NAMD}
L.~Kale, R.~Skeel, M.~Bhandarkar, R.~Brunner, A.~Gursoy, N.~Krawetz,
  J.~Phillips, A.~Shinozaki, K.~Varadarajan, and K.~Schulten.
\newblock {\em J. Comp. Phys.}, 151:283--312, 1999.

\bibitem{DuBostA01}
J.~P. DuBost, J.~M. Leger, J.~C. Colleter, P.~Levillain, and D.~Fompeydie.
\newblock {\em C. R. Acad. Sci. Paris S{\'e}r. II}, 292:965--968, 1981.

\bibitem{HolstA01}
M.~J. Holst, N.~A. Baker, and F.~Wang.
\newblock {\em J. Comput. Chem.}, 21:1343--1352, 2000.

\bibitem{PetersenA01}
M.~T. Neves-Petersen and S.~B. Petersen.
\newblock {\em Biotechnology Ann. Rev.}, 9:315--394, 2003.

\bibitem{HanggiA01}
P.~H{\"a}nggi, P.~Talkner, and M.~Borkovec.
\newblock {\em Rev. Mod. Phys.}, 62:251--341, 1990.

\bibitem{MummertA01}
M.~E. Mummert and E.~W.~Voss Jr.
\newblock {\em Biochemistry}, 35:8187--8192, 1996.

\bibitem{BoderA01}
E.~T. Boder, K.~S. Midelfort, and K.~D. Wittrup.
\newblock {\em Proc. Natl. Acad. Sci. U.S.A.}, 97:10701--10705, 2000.

\bibitem{HermannA01}
M.~E. Meyer-Hermann, P.~K. Maini, and D.~Iber.
\newblock {\em Math. Med. Biol.}, 23:255--277, 2006.

\end{thebibliography}

\end{document}